\documentclass[draft]{agujournal}
\usepackage{url}
\usepackage{amsmath}
\usepackage{graphicx}
\usepackage{float}
\journalname{Geophysical Research Letters}

\newcommand{\aap}{    {\it Astron. Astrophys.}}

\newcommand{\apj}{    {\it Astrophys. J.}}
\newcommand{\apjl}{   {\it Astrophys. J. Lett.}}

\newcommand{\solphys}{{\it Solar Phys.}}
 
\newcommand{\ssr}{    {\it Space Sci. Rev.}} 
\chardef\us=`\_

\begin{document}

%
%

\title{A new signal of the solar magnetic cycle: Opposite shifts of weak magnetic field distributions in the two hemispheres}

%
%

\authors{Tibebu Getachew\affil{1}, Ilpo Virtanen\affil{1}, and Kalevi Mursula\affil{1}}

\affiliation{1}{ReSoLVE Centre of Excellence, Space Climate Research Unit, University of Oulu, 90014 Oulu, Finland}


\correspondingauthor{Tibebu Getachew}{tibebu.ayalew@oulu.fi}


\begin{keypoints}
\item Shifts of weak photospheric magnetic field distributions in the two hemispheres follow the evolution of the trailing flux. 
\item  Hemispheric shifts change their sign in the late ascending or maximum phase of the solar cycle and attain their maximum in the early to mid-declining phase.	
\item  Shifts of the northern and southern hemisphere have opposite sign, but the southern hemisphere shifts are systematically larger in magnitude than in the north.
\item The hemispheric shifts have always the same sign as the polarity of the polar field in the respective hemisphere and solar cycle.
\item Evolution of hemispheric shifts gives a new signal of the solar magnetic cycle.
\end{keypoints}

%
%

\begin{abstract}
We study the asymmetric distribution of weak photospheric magnetic field values in the two hemispheres separately using synoptic maps from SDO/HMI, SOLIS/VSM and WSO during solar cycles 21-24. We calculate the weak-field asymmetry (shift) by fitting the distributions of weak-field values to a shifted Gaussian. Hemispheric shifts derived from the three data sets agree very well, and increase systematically when reducing the spatial resolution of the map. Shifts of the northern and southern hemisphere are typically opposite to each other. Shifts follow the evolution of the trailing flux and have a strong solar cycle variation with maxima in the early to mid-declining phase of the solar cycle. The sign of the hemispheric weak-field shift is always the same as the polarity of the polar field in the respective hemisphere and solar cycle. We also find that shifts in the south are systematically larger in absolute value than in the north.
	
\end{abstract}

\section{Introduction}
\label{introduction}
Sunspot cycle is a manifestation of the waxing and waning of intense magnetic field regions like sunspots on the solar surface. Since the magnetic polarity changes every 11 years \citep{Hale1919}, the solar magnetic cycle (the Hale cycle) is 22 years, consisting of two sunspot cycles \citep[for a historical review, see, e.g.,][]{Lidia2015,Stenflo2015}. While sunspot magnetic fields have been measured since the early $20^{th}$ century \citep{Hale1919}, the magnetic field of the whole photosphere has been measured only since 1950s \citep{Babcock1953}, with one of the longest consecutive series of measurements made at the Wilcox Solar Observatory since 1970s \citep{Svalgaard1978,Hoeksema1984,Hoeksema2010}.\\

Sunspots emerge at low to mid-latitudes of the solar surface, producing tilted bipolar magnetic regions (BMRs). The large-scale magnetic field observed at the solar photospheric surface is mainly formed by the evolution (diffusion, transport etc) of BMRs. According to Joy's law, the axis of the bipolar regions is tilted with respect to the east-west direction, the leading polarity being at a slightly lower latitude than the trailing one. The trailing polarity flux of the decaying BMRs forms poleward directed surges which cancel the old polar field and, eventually, create the new field of opposite polarity \citep{Babcock1961,Leighton1969}. Because the magnetic polarity of BMRs is opposite in the two hemispheres, the polar fields are also opposite to each other during most of the solar cycle. However, the evolution of surges is not identical in the two hemispheres. Moreover, the leading flux may also cause surges \citep[so called counter-surges,][]{Ulrich2013} which also affect the large-scale field at mid- and high latitudes.\\

The measured photospheric magnetic field consists mainly of rather weak fields than strong field values, irrespective of measurement accuracy. However, a small number of strong field values carry most of the total magnetic flux \citep{Stenflo1982}. Due to improved observational and diagnostic capabilities, there has been an increasing interest in the study of weak photospheric magnetic field values \citep[e.g.,][]{Lites2002,Berger2002,Berger2003,Stenflo2010,Stenflo2014}. We have recently shown \citep{Getachew2019} that the distribution of weak field values of the photospheric magnetic field is often asymmetric so that the distribution is slightly shifted toward either polarity. This shift was seen to be systematic in several simultaneous data sets. We argued that these weak-field shifts reflect a real asymmetry in the distribution of positive and negative weak-field values produced most effectively at the spatial scale of supergranulation \citep{Getachew2019}. \\

In this paper we study the distribution of weak field values and the related weak-field asymmetry (shift) separately for the northern and southern hemispheres using high-resolution data from the Helioseismic and Magnetic Imager (HMI) on-board the Solar Dynamics Observatory (SDO) and from the Synoptic Optical Long-term Investigations of the Sun (SOLIS) Vector SpectroMagnetograph (VSM), as well as low-resolution data from the Wilcox Solar Observatory (WSO). The paper is organized as follows. Section ~\ref{Data and methods} presents the data and methods used in this study. Section ~\ref{weak-field offset of HMI synoptic maps} studies on the hemispheric weak-field shifts in HMI data, and Section ~\ref{weak-field offset of VSM synoptic maps} the same for SOLIS/VSM data. Section ~\ref{weak-field offset of WSO synoptic maps} presents the long-term evolution of hemispheric shifts using WSO data. Finally, we discuss the results and give our conclusions in Section ~\ref{Discussion and conclusions}. \\
\section{Data and methods}
\label{Data and methods}

In this paper we use synoptic maps of the photospheric pseudo-radial (line-of-sight divided by cosine of latitude) magnetic field measured at HMI, SOLIS/VSM, and WSO to calculate the hemispheric weak-field asymmetry. HMI synoptic maps of the pseudo-radial field have a resolution of 3600*1440 pixels equally spaced in longitude and sine-latitude \citep{Liu2012,Hoeksema2014,Hayashi2015}. We note that the possible random zero-offsets of the high-resolution HMI magnetograms are removed when constructing these maps \citep [see ][]{Liu2012}. The HMI synoptic maps included in this paper cover Carrington rotations (CR) $2097-2214$, i.e., the time interval from 2010.4{\thinspace}--{\thinspace}2019.1. We derived four additional sets of medium- and low-resolution HMI synoptic maps (360*180, 180*72, 120*48, and 72*30) from the 3600*1440 synoptic maps using simple block averaging method \citep[see][]{Getachew2019} in order to investigate the effect of data resolution on the hemispheric weak-field asymmetries.\\

High-resolution synoptic maps of the (pseudo-radial) photospheric magnetic field have been produced at the National Solar Observatory (NSO) using SOLIS/VSM instrument for CR 2007{\thinspace}--{\thinspace}2196, i.e., between 2003.7{\thinspace}--{\thinspace}2017.8. The SOLIS/VSM synoptic maps used in this study have 1800*900 pixels equally spaced in longitude and sine-latitude, from which we calculated four sets of lower-resolution synoptic maps (360*180, 180*75, 120*50 and 72*30). VSM synoptic maps give the pseudo-radial fields without polar filling. \citep[Details on SOLIS/VSM data can be found, e.g., in][]{Bertello2014}. \\

The WSO synoptic maps (of the line-of-sight field) have a rather poor resolution of only 72*30 pixels in longitude and sine-latitude \citep{Hoeksema1984}. We calculated the pseudo-radial field from the line-of-sight magnetic fields. In this study we used all the $569$ published F-data (where missing data are filled by interpolation) WSO synoptic maps from CR 1642{\thinspace}--{\thinspace}2210, i.e., between 1976.3{\thinspace}--{\thinspace}2018.8. Note that in 1996-1999 and in 2001.1-2001.5 the WSO data are erroneous and in clear disagreement with other data \citep[see, e.g.,][]{Virtanen2016,Virtanen2017}. This period of erroneous data is left out from the analysis of this paper. The WSO measurements are good in that the same instrumentation has been in operation since measurements have started. \citep[Details about the WSO magnetograph, data calibration and reduction methods can be found, e.g., in][]{Hoeksema1984}. \\ 

We calculate the weak-field shifts in the same way as in \cite{Getachew2019}. The histogram distribution of weak field values \citep[here within $\pm 10G$; note that results are insensitive to the range, as discussed by][]{Getachew2019} for each synoptic map, separately for northern and southern hemispheres, is fitted with a shifted Gaussian:

\begin{equation}
\phi =c\exp\left({-\frac{1}{2}\left(\frac{B_{i}-a}{b}\right)^{2}}\right),
\label{equ:Gaussian}
\end{equation}
where $B_{i}$ are the observed (weak) field values, and the three fit parameters $a$, $b$ and $c$ give the position of the peak (the weak-field shift), the width of the Gaussian distribution (standard deviation) and the amplitude, respectively.

\section{Hemispheric Weak-field shifts of HMI}
\label{weak-field offset of HMI synoptic maps}

\begin{figure}[!t] 
	\centering
	\includegraphics[width=\columnwidth]{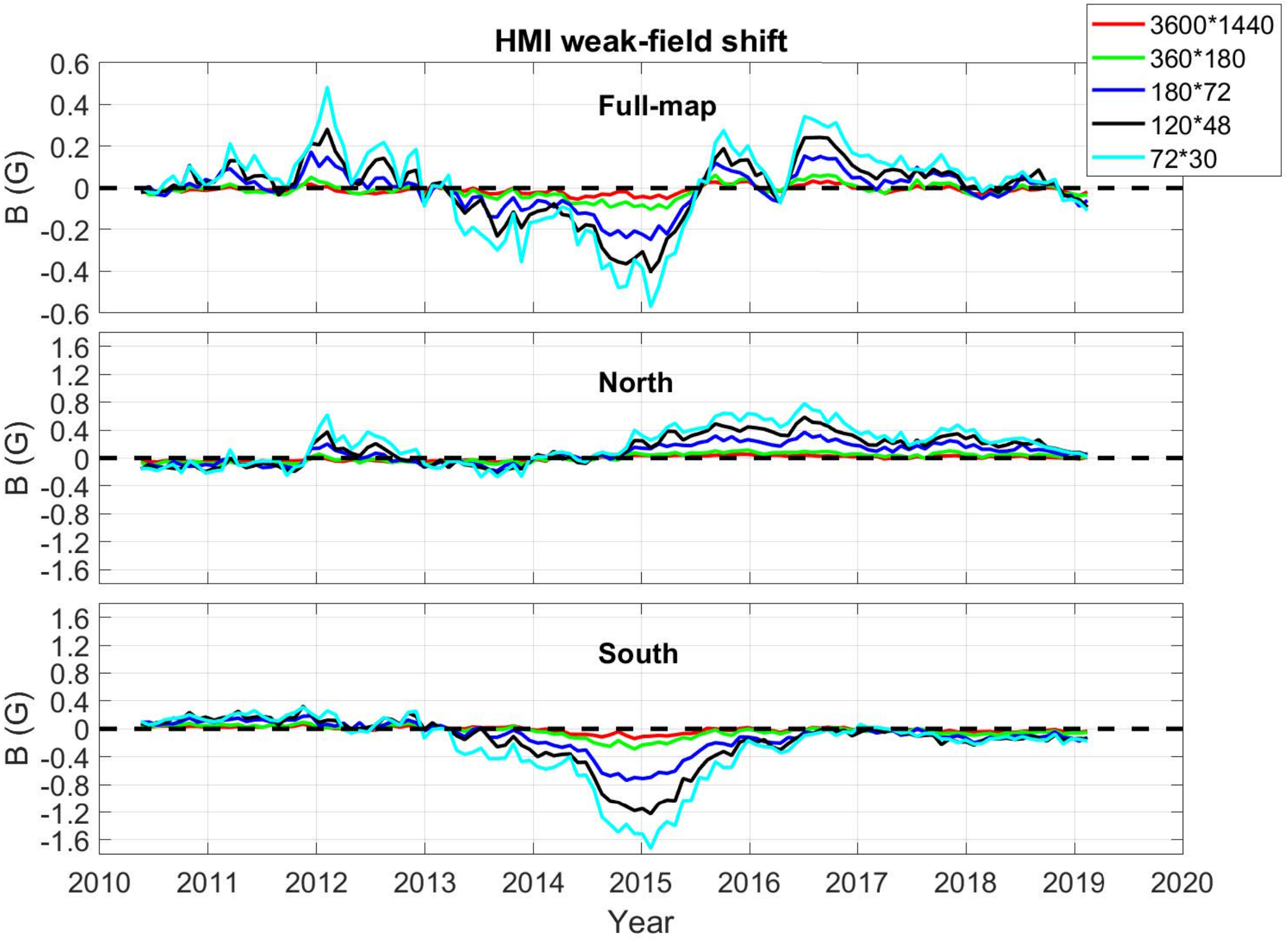}
	\caption{Weak-field shifts of HMI synoptic maps of 3600*1440 (red line), 360*180 (green line), 180*72 (blue line), 120*48 (black line) and 72*30 (cyan line) resolution. Top panel uses data from full synoptic maps, while the middle and bottom panels use data only from the northern and southern hemispheres, respectively.}
	\label{fig:hmi_weak_field_offset_per_hem}
\end{figure}

Figure \ref{fig:hmi_weak_field_offset_per_hem} shows the HMI weak-field shift values for each synoptic map at the five resolutions mentioned above. The top panel presents the shift values using the full synoptic maps \cite[cf. Figure 3 in ][]{Getachew2019}, and the middle and bottom panels show the shifts using only the northern and southern hemisphere of the synoptic map, respectively. Top panel shows that the full-map shifts mainly occur at the same time and have the same sign (positive or negative) for all five resolutions. The same is true for the hemispheric weak-field shifts in the two other panels of Figure \ref{fig:hmi_weak_field_offset_per_hem}. Also, for all three cases (full-map, northern and southern hemisphere) the weak-field shifts increase with decreasing resolution of the maps, as already found for full maps in \cite{Getachew2019}. Shifts increase significantly when the resolution decreases from 360*180 to 180*72, reflecting the approach to the supergranulation scale of the photospheric field \citep{Getachew2019}. Note that the full-map shifts are typically quite close to the sum of the northern and southern shifts of the respective rotation. However, this is not always exactly the case, because the hemispheric distributions may be occasionally quite different. \\

Figure \ref{fig:hmi_weak_field_offset_per_hem} (middle panel) shows that the northern hemisphere has a weak negative shift in 2010{\thinspace}--{\thinspace}2011, a clearly larger, positive shift in 2012 and then again a weak, mainly negative shift in 2013. Since 2014, until the end of data in 2019, the northern hemisphere has a fairly large, systematically positive weak-field shift. \\

The southern hemisphere (Figure \ref{fig:hmi_weak_field_offset_per_hem} bottom panel) has a rather weak, mostly positive shift in 2010{\thinspace}--{\thinspace}2012. From 2013 onwards, the southern shift is mostly negative and attains very large values in 2014{\thinspace}--{\thinspace}2015. Accordingly, the shifts in the northern and southern hemisphere were oppositely directed for most of the HMI period. Out of 118 rotations included in Figure \ref{fig:hmi_weak_field_offset_per_hem}, the northern and southern hemisphere have oppositely oriented shifts at least in 87 rotations ($74\%$). Thus there is a clear tendency for the simultaneous shifts to be oppositely oriented in the northern and southern hemisphere.\\

\begin{table*}[!t]
	\caption{Mean values of HMI hemispheric weak-field shifts (in Gauss) using all shifts and dominantly signed shifts in each hemisphere. The upper and lower confidence intervals are given in parenthesis.}
	\resizebox{\columnwidth}{!}{%

		\centering
		
		\begin{tabular}{|c|c|c|c|c|c|c|}
			\hline
			Hemisphere&	mean value of & 3600*1440&360*180&180*72&120*48& 72*30\\ 
			\hline  		
			
			&	all shifts & \shortstack{ -0.007 \\(-0.013, -0.0004)} & \shortstack{ 0.004 \\(-0.007, 0.015)} &  \shortstack{ 0.066 \\( 0.041, 0.091)}  &   \shortstack{ 0.122\\ (0.084,  0.160)}& \shortstack{ 0.184 \\(0.136, 0.233)}\\
			\cline{2-7}
			North&positive shifts& \shortstack{ 0.027\\ (0.023, 0.031)}  &  \shortstack{ 0.056 \\(0.049, 0.063)}  &\shortstack{  0.150\\ (0.130, 0.169)}  & \shortstack{ 0.242 \\(0.210, 0.274)}  &  \shortstack{ 0.320 \\(0.280, 0.363)}\\
			\hline
			&	all shifts&\shortstack{-0.013 \\(-0.021, -0.005)}& \shortstack{-0.032 \\(-0.047, -0.018) } & \shortstack{-0.107 \\(-0.147, -0.068)}& \shortstack{-0.171\\ (-0.234, -0.108)}  & \shortstack{-0.244\\ (-0.330, -0.161)}\\
			\cline{2-7}
			\cline{2-7}
			South&	negative shifts&\shortstack{ -0.042 \\(-0.050, -0.034)}  & \shortstack{-0.074\\ (-0.090, -0.058)}  & \shortstack{ -0.204\\ ( -0.249, -0.159)}  & \shortstack{ -0.320 \\(-0.395, -0.244)}& \shortstack{-0.420 \\(-0.514, -0.317)}\\
			\hline
		\end{tabular}

	}
	
	\label{table:Mean values of hmi hemispheric shifts}
	
\end{table*}

Table \ref{table:Mean values of hmi hemispheric shifts} gives the mean values of HMI hemispheric shifts with lower and upper limits at 95\% given in parenthesis. In order to calculate the lower and upper confidence intervals (CI) of the mean values of shifts, we applied the Student's t-test given as

 \begin{equation}
 \begin{split}
CI=\frac{1}{n}\sum_{i=1}^{n}\mu_{i}\pm S_{\mu} t_{\alpha,n-1}\\
\end{split}
\label{equ:t_test}
\end{equation}
where $S_{\mu}$ is the error of the mean of individual shifts $\mu_{i}$, and $t_{\alpha,n-1}$ is the precalculated statistic value for the t-distribution , where $\alpha$ is the significance level (for a $95\%$ confidence interval, $\alpha=0.025$) and  $n$ is the number of shift values ($n= 118$ for all shifts of HMI data).\\

We have calculated the means of all shifts  over the full time interval, and also the means of the dominantly signed shifts in either hemispheres, i.e., positive in the north and negative in the south. Table \ref{table:Mean values of hmi hemispheric shifts} shows that the positive shifts in the north and the negative shifts in the south increase (in absolute value) systematically with reducing spatial resolution, from about $0.03$ ($-0.04$) in the 3600*1440 maps to $0.32$ ($-0.42$) in the 72*30 maps in the north (south, respectively). The largest step in this decadal increase occurs between 360*180 and 180*72, where shift increases roughly by a factor of $2.5$ to $3$. Note also that the negative shifts in the south are systematically (for all resolutions) larger in absolute value than the positive shifts in the north.\\

\section{Hemispheric Weak-field shifts of SOLIS/VSM}
\label{weak-field offset of VSM synoptic maps}

\begin{figure}[!t] 
	\centering
	\includegraphics[width=\columnwidth]{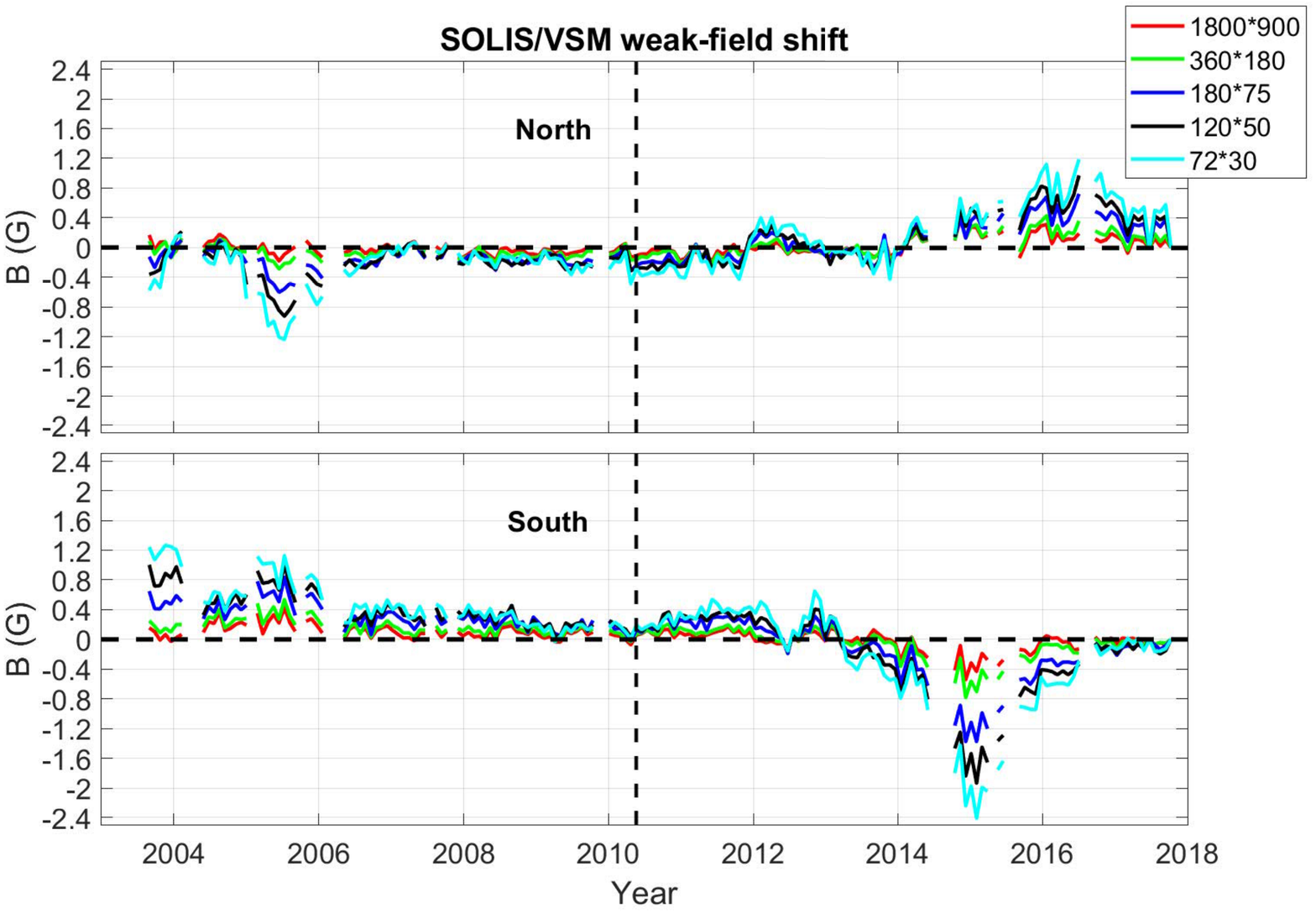}
	\caption{Hemispheric weak-field shifts of SOLIS/VSM synoptic map for 1800*900 (red line), 360*180 (green line), 180*75 (blue line), 120*50 (black line) and 72*30 (cyan line) resolutions. Upper panel is for the northern hemisphere and lower panel is for the southern hemisphere. The start time of HMI data is marked by dashed vertical line.}
	\label{fig:vsm_weak_field_offset_per_hem}
\end{figure}

Figure \ref{fig:vsm_weak_field_offset_per_hem} shows the hemispheric weak-field shifts for SOLIS/VSM synoptic maps at the five resolutions of 1800*900, 360*180, 180*75, 120*50 and 72*30  in 2003.7{\thinspace}--{\thinspace}2017.8. Comparing Figures \ref{fig:hmi_weak_field_offset_per_hem} and \ref{fig:vsm_weak_field_offset_per_hem}, in most of the simultaneous rotations, the sign of both the northern and southern hemisphere shifts are the same in HMI and SOLIS/VSM. Out of the 100 rotations of the overlapping time, $74$ ($87$) HMI shifts of 360*180 (72*30 maps, respectively) maps have the same sign as SOLIS/VSM maps in the northern hemisphere, and $77$ ($82$) shifts in the southern hemisphere. This gives strong evidence for the validity and universality of the obtained hemispheric shifts.\\

We have also calculated the correlation coefficients ($r$) and the corresponding p-values between SOLIS/VSM and HMI shifts in the northern and southern hemispheres for all possible resolution combinations (5 SOLIS/VSM $\times$ 5 HMI). For all $25$ combinations and for both hemispheres, the correlations are high and extremely significant. Correlations were overall slightly more important in the south than in the north. The lowest correlation was found in the northern hemisphere between the highest SOLIS/VSM and lowest HMI resolution ($r=0.6$, $p<10^{-9}$) and the highest correlation in the southern hemisphere for the lowest resolution of both data sets ($r=0.96$, $p<10^{-51}$).\\

Figure \ref{fig:vsm_weak_field_offset_per_hem} shows that the SOLIS/VSM shifts are mostly negative in the northern hemisphere in 2003{\thinspace}--{\thinspace}2011, and mostly positive thereafter (cf. Figure \ref{fig:hmi_weak_field_offset_per_hem}). The largest negative shifts in the north are found in 2005. SOLIS/VSM shifts in the southern hemisphere are mostly positive in 2003{\thinspace}--{\thinspace}2012, and mostly negative thereafter. The highest positive shifts in the south are found in 2004{\thinspace}--{\thinspace}2005. These results further strengthen the overall anti-correlation between the northern and southern shifts, as indicated earlier by HMI results. They also suggest that there is a change in the sign of the hemispheric shifts from one solar cycle to the next.\\

The mean SOLIS/VSM shifts are systematically larger than the HMI shifts for the same resolution and hemisphere (note the different scales of y-axis in Figures \ref{fig:hmi_weak_field_offset_per_hem} and \ref{fig:vsm_weak_field_offset_per_hem}). The SOLIS/HMI ratios vary from a factor of about two for 360*180 resolution to only about 30-50\% for 72*30 resolution. The hemispheric shifts of SOLIS/VSM increase systematically when the resolution decreases, as already seen for HMI. Also similarly with HMI, the SOLIS/VSM mean shifts are larger in the south than in the north.

\section{Hemispheric Weak-field shifts of WSO}
\label{weak-field offset of WSO synoptic maps}

\begin{figure}[!t] 
	\centering
	\includegraphics[width=\columnwidth]{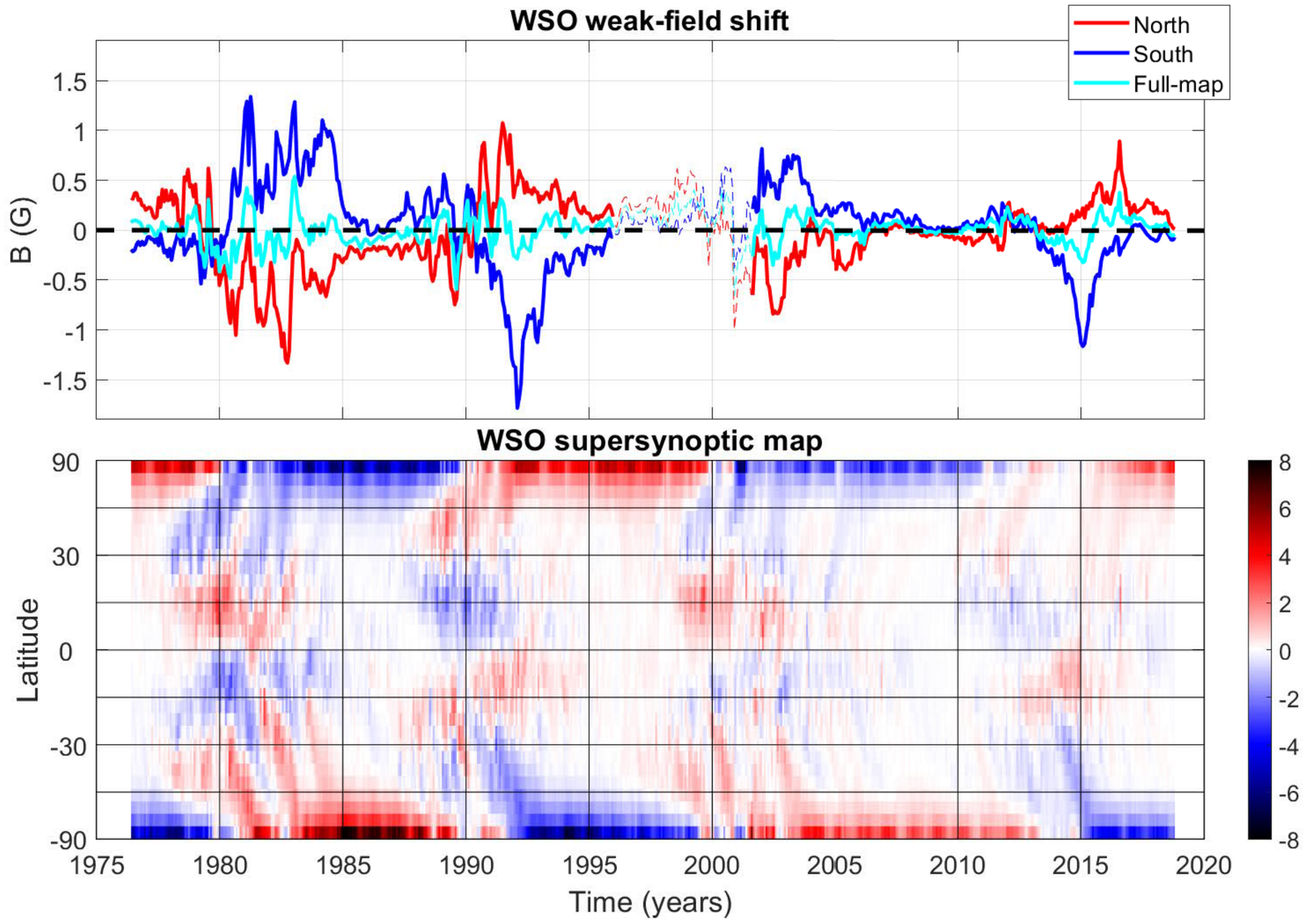}
	\caption{Top panel shows WSO synoptic map weak-field shifts separately for the northern (red line), southern (blue line) hemispheres and the full-disk maps (cyan line). Weak-field asymmetries during the erroneous data period (1996{\thinspace}--{\thinspace}2001) are shown by a dashed line. Bottom panel shows photospheric magnetic field butterfly diagram (super-synoptic map) from WSO synoptic maps. In the bottom panel plot, red color indicates positive (outward) polarity and blue negative (inward) polarity. Units are in Gauss.}
	\label{fig:wso_weak_field_offset_per_hem}
\end{figure}

Top panel of Figure \ref{fig:wso_weak_field_offset_per_hem} shows the full-map and hemispheric weak-field shifts for WSO in CR 1642{\thinspace}--{\thinspace}2210, i.e., between 1976.3{\thinspace}--{\thinspace}2018.8, including four solar cycles. Figure \ref{fig:wso_weak_field_offset_per_hem} (top panel) verifies that the northern and southern shifts evolve roughly in anti-correlation, and depict a clear variation over the solar cycle. Hemispheric shifts have  maxima in the early to mid-declining phase of the solar cycle, and decrease to very small absolute values until the solar minimum, but change their sign typically only in the late ascending to maximum phase of the solar cycle.\\

Bottom panel of Figure \ref{fig:wso_weak_field_offset_per_hem} shows the map of longitudinally averaged magnetic field derived from WSO synoptic maps, the so called photospheric super-synoptic map, also called the magnetic field butterfly diagram. This diagram depicts the overall pattern of the solar magnetic cycle, i.e., the alteration of magnetic polarity according to the 22-year cycle at the solar surface. It also shows the transport of surges of new flux to the poles and the eventual reversal of the polar fields.\\

Figure \ref{fig:wso_weak_field_offset_per_hem} suggests that hemispheric weak-field shifts are related to the poleward surges of magnetic flux and the evolution of polar fields. The different timing of surges in the north and south (see, e.g., 1990, 1991) is also seen in the different time evolution of the northern and southern hemisphere weak-field shifts.\\

Table \ref{table:Mean_values_of_wso_hemispheric_shift} presents the mean shift for each hemisphere and each solar cycle, separately. The times were determined from the change of the sign of the shift in the corresponding hemisphere.  Table \ref{table:Mean_values_of_wso_hemispheric_shift} shows clearly that the mean shifts follow the sign (polarity) of the solar pole in that hemisphere and cycle. Moreover, in each cycle (and overall) the southern hemisphere shifts are somewhat larger than in the northern hemisphere. This difference is particularly strong during the two even cycles where the shift  ratios are about 1.42 and 1.47 for solar cycle 22 and solar cycle 24, while they are only about 1.1 and 1.05 for the two odd cycles.\\

\begin{table}[h]
	\caption{Mean values of WSO hemispheric weak-field shifts in Gauss for each solar cycle (SC). The upper and lower confidence intervals are given in  parenthesis.}
	\resizebox{\columnwidth}{!}{%
		\centering
		
		\begin{tabular}{|c|c|c|c|c|c|c|c|c|}
			\hline
			Hemisphere&	& SC21&SC22&SC23&SC24\\ 
			\hline
			&times&1979.6{\thinspace}--{\thinspace}1990.3 &1990.3{\thinspace}--{\thinspace}1996 &2001{\thinspace}--{\thinspace}2011.9 &2011.9{\thinspace}--{\thinspace}2018.8\\			
			\cline{2-6}
			North	&all shifts&-0.388 (-0.437, -0.339)&0.367 (0.313, 0.421)&-0.147 (-0.182, -0.112)& 0.196 (0.161, 0.230)\\
			\cline{2-6}
			&dominant shift&-0.403 (-0.451, -0.356)&0.367 ( 0.313, 0.421)&-0.193 (-0.233, -0.154)&0.213 (0.179, 0.248)\\			
			\hline
			&time&1980.4{\thinspace}--{\thinspace}1990.4&1990.4{\thinspace}--{\thinspace}1996&2001{\thinspace}--{\thinspace}2013.2&2013.9{\thinspace}--{\thinspace}2018.8\\
			\cline{2-6}
			South&all shifts&0.396 (0.334, 0.459)&-0.516 (-0.612, -0.419)&0.196 (0.166, 0.227)&-0.272 (-0.343, -0.201)\\ 
			\cline{2-6}
			&dominant shift&0.444 (0.382, 0.506)&-0.523 (-0.620, -0.426)&0.202 (0.171, 0.233)&-0.314 (-0.388, -0.239)\\ 
			\hline
		\end{tabular}
		}
	
	\label{table:Mean_values_of_wso_hemispheric_shift}
	
\end{table}

The WSO hemispheric shifts agree very well  with the signs of the corresponding SOLIS/VSM and HMI shifts, and the absolute values of the WSO shifts are typically between those of HMI and SOLIS/VSM. Out of the $169$ overlapping rotations, $117$, $126$, $133$, $139$ and $140$ SOLIS/VSM (1800*900, 360*180, 180*75, 120*50 and 72*30) shifts had the same sign in the north as WSO, and $136$, $151$, $158$, $157$ and $157$ shifts in the south. Correlation in hemispheric shifts between WSO and SOLIS/VSM are all high and statistically significant, at least with 
$r=0.5$ ($p<10^{-10}$). Correlations are slightly better for the south than north, as found earlier between SOLIS/VSM and HMI.\\

\section{Discussion and conclusions}
\label{Discussion and conclusions}
In this paper we have studied the hemispheric shifts of weak photospheric magnetic field values using HMI, SOLIS/VSM and WSO data sets at different resolutions. Hemispheric shifts were calculated by fitting the histogram distribution of hemispheric weak-field values (here within $\pm 10G$) for each synoptic map to a shifted Gaussian \citep{Getachew2019}.\\

As found earlier for the full-map shifts \citep{Getachew2019}, the hemispheric shifts have systematically the same sign for different map resolutions, and increase with reduced spatial resolution. The shifts increase significantly in both hemispheres when the resolution decreases from 360*180 to 180*72, reflecting the approach to the supergranulation scale of the photospheric field \citep{Getachew2019}. Moreover, the shifts have mainly the same sign for simultaneous rotations for all the three independent instruments (HMI, SOLIS/VSM, WSO) in each hemisphere. The absolute values of the shifts are somewhat different, with SOLIS/VSM shifts being largest and HMI smallest, but this difference decreases when reducing the map resolution.\\

We find that the northern and southern hemisphere shifts are typically opposite to each other, but the simultaneous shifts do not often have the same (absolute) magnitude. These differences in the magnitudes and time evolutions of the hemispheric shifts explain the full-map shifts studied 
earlier in \citet{Getachew2019}.\\

Hemispheric shifts depict a clear evolution over the solar cycle. They change their sign in the late ascending or maximum phase of the cycle and attain their maximum in the early to mid-declining phase. The hemispheric shifts have always the same sign as the polarity of the respective pole of the future solar minimum. Thus the hemispheric shifts have a relation to the evolution of the polar fields.\\

We found that the southern hemisphere shifts are systematically larger than in the north in all the three data sets and solar cycles. This is particularly true for the even solar cycles (SC22 and SC24), where the southern shifts are roughly 50\% larger at WSO resolution than the northern shifts. This difference may contribute to the earlier observed stronger field intensity in the south and the related southward shift of the heliospheric current sheet, also called the bashful ballerina phenomenon \citep{Mursula2003,Mursula2004,Zhao2005,Hiltula2006,Virtanen2010,Virtanen2014,Virtanen2016}.
\newpage
\acknowledgments

 We acknowledge the financial support by the Academy of Finland to the ReSoLVE Centre of Excellence (project no. 307411). Wilcox Solar Observatory data used in this study was obtained via the web site  \url{http://wso.stanford.edu} courtesy of J.T. Hoeksema. Data were acquired by SOLIS instruments operated by NISP/NSO/AURA/NSF. HMI data are courtesy of the Joint Science Operations Center (JSOC) Science Data Processing team at Stanford University. Data used in this study was obtained from the following web sites: WSO: \url{http://wso.stanford.edu} SOLIS: \url{http://solis.nso.edu/0/vsm/vsm_maps.php} HMI: \url{http://jsoc.stanford.edu/data/hmi/synoptic}\\



\end{document}